# Nonequivalent Ensembles for the Mean-Field $\varphi^6$ Spin Model

**Abstract** We derive the thermodynamic entropy of the mean field $\varphi^6$ spin model in the framework of the micro-canonical ensemble as a function of the energy and magnetization. Using the theory of large deviations and Rugh's micro-canonical formalism we obtain the entropy and its derivatives and study the thermodynamic properties of $\varphi^6$ spin model. The interesting point we found is that like $\varphi^4$ model the entropy is a concave function of the energy for all values of the magnetization, but is non-concave as a function of the magnetization for some values of the energy. This means that the magnetic susceptibility of the model can be negative for some values of the energy and magnetization in the micro-canonical formalism. This leads to the inequivalence of the micro-canonical and canonical ensembles. It is also shown that this mean-field model , displays a first-order phase transition due to the magnetic field. Finally we compare the results of the mean-field $\varphi^6$ and $\varphi^4$ spin models.

**Keywords** Statistical Mechanics, Mean-Field $\varphi^6$ Spin Models, Ensemble Inequivalence , Non-Concave Entropies

## 1. Introduction

The micro-canonical and canonical ensembles are the two main probability distributions with respect to which the equilibrium properties of statistical mechanical models are calculated. Despite the two ensembles model two different physical situations, it is generally believed that the ensembles give equivalent results in the thermodynamic limit; i.e., in the limit in which the volume of the system tends to infinity. The equivalence of the micro-canonical and canonical ensembles is most usually explained by saying that although the canonical ensemble is not a fixed-mean-energy ensemble like the micro-canonical ensemble, it must 'converge' to a fixed mean energy ensemble in the thermodynamic limit ,and so must become or must realize a micro-canonical ensemble in that limit .This convergence can be proved to hold for non-interacting systems such as the perfect gas and for a variety of weakly interacting systems. For general systems, however this convergence is not valid. In fact, in the past three and a half decades, numerous statistical models have been discovered having micro-canonical equilibrium properties that cannot be accounted for within the framework of the canonical ensemble. For systems with short range interactions, the choice of the statistical ensemble is typically of minor importance and could be considered a finite size effect: differences between, say micro-canonical and canonical expectation values are known to vanish in the thermodynamic limit of large system size and the various statistical ensembles become equivalent . In the presence of long-range interactions this is in general not the case, and micro-canonical and canonical approaches can lead to different thermodynamic properties even in the infinite system limit . In the astrophysical context, nonequivalence of ensembles and the importance of micro-canonical calculations have long been known for gravitational systems. Although equivalence of ensembles had been proven only for short range interactions, it was tacitly assumed by most physicists to hold in general. Therefore it came as a surprise to many that equivalence does not necessarily hold for long range systems, in particular in the presence of a discontinuous phase transition. Until now, the phenomenon of nonequivalent ensembles has been identified and analyzed almost exclusively by determining regions of the mean energy where the micro-canonical entropy function is anomalously non-concave or by determining regions of the mean energy where the heat capacity, calculated micro-canonically is negative. The existence of such non-concave anomalies invalidates yet another tacit principle of statistical mechanics which states that the one should always be able to express the micro-canonical entropy, the basic thermodynamic function for the micro-canonical ensemble, as the Legendre transform of the free energy, the basic thermodynamic function for the canonical ensemble. Indeed, if the micro-canonical entropy is to be expressed as the

Legendre transform of the canonical free energy, then the former function must necessarily be concave on its domain of definition. More recently non-concave anomalies in the micro-canonical entropy as well as negative heat capacities have been observed in models of fluid turbulence and models of plasmas, in addition to long-range and mean-field spin models, including the mean field XY model and the mean field Blume-Emery-Griffiths (BEG) model.

One of the motivations for the study of $\varphi^6$ models is the search for soliton. Another reason for interest is that $\varphi^6$ models are the simplest systems with continuous variables that exhibit a rich phase diagram, with first and second order phase transitions and a tricritical point. This structure was observed in the study of the quantum mechanics of a single site, three well potential when classical, perturbative and mean field arguments were used and bubble solutions, their relation to the phase transitions and the question of their stability, both relativistically and non-relativistically. In recent years the $\varphi^6$ model and its application in different physical systems including the following problems have been studied extensively: the crossover from a quantum $\varphi^6$ theory to a renormalized two-dimensional classical nonlinear sigma model [1], alpha matter on a lattice [2], first-order electroweak phase transition (EWPT) due to a dimension six operator in the effective Higgs potential [3], first order phase transitions in confined systems [4], effective potential and spontaneous symmetry breaking in the non-commutative $\varphi^6$ model [5], bubble dynamics in quantum phase transitions [6], the canonical transformation and duality in the $\varphi^6$ theory [7], hermitian matrix model $\varphi^6$ for 2D quantum gravity [8], phase structure of the generalized two dimensional Yang-Mills theories on sphere [9], tricritical Ising model near criticality [10], spontaneous symmetry breaking at two Loop in 3D massless scalar electrodynamics [11], Ising model in the ferromagnetic phase [12], statistical mechanics of nonlinear coherent structures and kinks in the $\varphi^6$ model [13], growth kinetics in the $\varphi^6$ N-Component model [14], stability of Q-balls [15], the liquid states of pion condensate and hot pion matter [16], instantons and conformal holography [17], first order phase transitions in superconducting films [18], field theoretic description of ionic crystallization in the restricted primitive mode [19].

This increasing interest in $\varphi^6$ model and its many application in physics, motivated us to study the statistical mechanics of $\varphi^6$ spin models. It is worth to mention that the $\varphi^4$ spin model has been studied in [20,21].

## 2. The Mean Field $\varphi^6$ Model and its Thermodynamic

The Hamiltonian of the mean-field $\varphi^6$ model is given by the following expression :

$$H = \sum_{i=1}^{N} \frac{p_i^2}{2} - \frac{q_i^2}{4} + \frac{q_i^4}{4} + \frac{q_i^6}{6} - \frac{1}{4N}\sum_{i,j=1}^{N} q_i q_j \tag{1}$$

where $q_i$ and $p_i$ are the canonical coordinates of unit mass. The entropy of the system in terms of its mean energy and magnetization is defined as (we set the Boltzman constant $\kappa_b = 1$) :

$$S(\epsilon, m) = \lim_{N \to \infty} \frac{1}{N} \ln \int \delta(\epsilon(x) - \epsilon) \delta(m(x) - m) \, dx \tag{2}$$

$m(x)$ is the mean magnetization $Nm(x) = M$, where $N$ is the number of particles within the system and $x$ represents a point in the phase space. The relationship between magnetization $M$ and the canonical coordinates is given by $m(x) = \frac{1}{N}\sum_{i=1}^{N} q_i$.

If $S(\epsilon,m)$ were concave, then one would calculate it as the Legendre transform of the canonical free energy. Actually for concave entropies, the free energy and the entropy are the Legendre transform of one another [22] :

$$S(\varepsilon) = \varphi^*(\varepsilon) = \inf_\beta\{\beta\varepsilon - \varphi(\beta)\} \tag{3}$$

where :

$$\varphi(\beta) = -\lim_{N \to \infty} \frac{1}{N} \ln Z_N(\beta) \tag{4}$$

is the free energy and $Z_N(\beta) = \int \exp(-\beta H(\omega)) \, d\omega$ is the partition function of the system and $\beta = \frac{1}{T}$. For non-concave entropies we use two methods : i.e., a) method of large deviation; b) Rugh's method, and compare the results.

## 3. Calculation of the Entropy Using Large Deviation Method

It is convenient first to introduce the large deviation principle [23,24]. Suppose $A_n$ is a random variable (n is an integer) and let $P(A_n \in A)$ is the probability that $A_n$ takes on a value in a set A, then the large deviation principle is satisfied by $P$ if the following limit exists [23] :

$$\lim_{n \to \infty} -\frac{1}{n} \ln P(A_n \in A) = I(a) \tag{5}$$

where $I(a)$ is the rate function.

**Gärtner-Ellis Theorem :** The scaled cumulant generating function of $A_n$ is defined by :

$$\lambda(k) = \lim_{n\to\infty} \frac{1}{n} \ln\langle \exp(nkA_n)\rangle \tag{6}$$

Gärtner-Ellis theorem [22,23,25] states that, if $\lambda(k)$ exists and is differentiable for all $k \in R$, then $A_n$ satisfies a large deviation principle with the rate function I(a) which is the Legendre-Fenchel transform of $\lambda(k)$ and is given by $I(a) = sup_k\{ka - \lambda(k)\}$.

**Varadhan's Theorem :** If $A_n$ satisfies a large deviation principle with rate function $I(a)$, then λ(f) is the Legendre-fenchel transform of $I(a)$ [23,26]:

$$\lambda(f) = \lim_{n\to\infty} \ln\langle e^{nf(A_n)}\rangle = sup_a\{f(a) - I(a)\} \tag{7}$$

where $f$ is an arbitrary continuous function of $A_n$. This theorem implies that for non-concave entropies, the entropy and the free energy are the Legendre-Fenchel transform of one another. Therefore the micro-canonical large deviation principle allows one to calculate the entropy in terms of a maximization involving a macro-state entropy. In general for the entropy of the system we have :

$$S(u) = Sup_\eta \, \tilde{S}(u, \eta) \tag{8}$$

where η represents a set of macro variables that the quantities such as the mean energy $\epsilon(x)$ and the mean magnetization m(x) can be considered as functions of them :

$$S(\epsilon, m) = \underset{\eta: \tilde{\epsilon}(\eta) = \epsilon, \tilde{m}(\eta) = m}{Sup} \tilde{S}(\eta) \tag{9}$$

where :

$$\tilde{S}(\eta) = \lim \frac{1}{N} \ln \int \delta(\eta(x) - \eta) \, dx \tag{10}$$

$$\epsilon(x) = \tilde{\epsilon}(\eta(x)) \quad , \quad m(x) = \tilde{m}(\eta(x)) \tag{11}$$

$\tilde{S}(\eta)$ is given by Eqs.(3) and (4) with the following replacements : $\beta u \to \lambda.\eta$ and $Z_N(\beta) \to \int \exp(-N\lambda.\eta(x))\, dx$.

For $\Phi^6$ model an appropriate choice of macro state is the vector $\mu=(m, k, v)$, where m, k and $v$ are the mean magnetization, the mean kinetic energy and the mean potential energy respectively:

$$k = \frac{1}{2N} \sum p_i^2 \tag{12}$$

$$v = \frac{1}{N} \sum \left( \frac{q_i^6}{6} + \frac{q_i^4 - q_i^2}{4} \right) \tag{13}$$

So we can write $\tilde{\epsilon}(\eta)$ as follows:

$$\tilde{\epsilon}(\eta) = k + v - \frac{m^2}{4} \tag{14}$$

One can calculate the entropy $\tilde{S}(\eta)$ using the legendre transformation. It is shown in [20] that the $S(\epsilon, m)$ is given by the following expression:

$$S(\epsilon, m) = Sup_v \left\{ \frac{1}{2} \ln \left[ \epsilon - v + \frac{m^2}{4} \right] + \frac{1}{2} \ln 4\pi e + \tilde{S}(m, v) \right\} \tag{15}$$

## 4. Rugh's Formalism for the Mean-Field $\varphi^6$ Model

We denote the micro-canonical ensemble average (μ-average) and the Liouville volume by $\langle \Phi; E \rangle$ and $d\omega = d^d x d^d p$ respectively. The μ-average of an observable Φ is given by the following expression :

$$\langle \Phi; E \rangle = \frac{\mu_E(\Phi)}{\mu_E(1)} \; ; \quad \mu_E(\Phi) = \int d\omega \, \delta(H - E) \, \Phi \tag{16}$$

The μ-entropy and the inverse temperature are as follows :

$$S(E) = \ln \mu_E(1) \; ; \quad \frac{1}{T(E)} = \frac{\partial S(E)}{\partial E} \tag{17}$$

According to a theorem proven in [27], if $\Sigma_E$ is a regular energy surface of the Hamiltonian H and $X_i$ is a vector field defined in a neighbourhood of $\Sigma_E$ satisfying dH(X)≡ 1, then the energy derivative of the μ-average of the observable Φ

is given by :

$$\frac{\partial}{\partial E}\langle \Phi; E\rangle = \langle \nabla \cdot (\Phi X); E\rangle - \frac{\langle \Phi; E\rangle}{T(E)} \quad (18)$$

In the same way it is shown in [28], that :

$$\frac{\partial}{\partial \lambda_k}\langle \Phi\rangle_\mu = -\langle \nabla \cdot (\Phi \frac{\partial H_\lambda}{\partial \lambda_k} X)\rangle_\mu - P_\mu^k \langle \Phi\rangle_\mu \quad (19)$$

where $\langle \Phi\rangle_\mu = \langle \Phi; E, \lambda\rangle$ is the average of $\Phi$ in the micro-canonical ensemble and $\lambda$ is a set of parameters which the Hamiltonian can depend on them, for instance the magnetization of the system. $P_\mu^k = \frac{\partial S_\mu}{\partial \lambda_k} = -\langle \nabla \cdot (\frac{\partial H_\lambda}{\partial \lambda_k} X)\rangle$ is the 'generalized' pressure.
Form Eqs.(18) and (19) for the derivatives of the entropy we have :

$$\frac{\partial}{\partial E} S(E, M) = \langle \nabla \cdot X\rangle_{E,M} = \frac{1}{T(E,M)} \quad (20)$$

and:

$$\frac{\partial}{\partial M} S(E, M) = -\langle \nabla \cdot \left(\frac{\partial H}{\partial M} X\right)\rangle_{E,M} \quad (21)$$

By introducing the magnetization constraint into the Hamiltonian H(x) one can obtain the Hamiltonian H(x;M). To do this we eliminate one of the $q_i$, for example $q_N = M - \sum_{i=1}^{N-1} q_i$. For the mean field $\varphi^6$ model we have :

$$H(x; M) = \sum_{i=1}^{N-1} \frac{p_i^2}{2} - \frac{1}{2N}\sum_{i,j}^{1,N-1} q_i q_j - \sum_{i=1}^{N-1}\left(\frac{q_i^2 - q_i^4}{4} - \frac{1}{6}q_i^6\right) - \frac{1}{4}\left(M - \sum_{i=1}^{N-1} q_i\right)^2 + \frac{1}{4}\left(M - \sum_{i=1}^{N-1} q_i\right)^4$$

$$+ \frac{1}{6}(M - \sum_{i=1}^{N-1} q_i)^6 - \frac{M^2}{4N} \quad (22)$$

For the vector field X, we follow Rough [27] and choose : $X = \frac{1}{2K_c}(p_1, \ldots, p_{N-1}, 0, \ldots, 0)$ where $K_c$ is the kinetic part of H(x ;M). From Eq.(20) we have:

$$\frac{1}{T(E,M)} = \langle \frac{N-3}{2K_c}\rangle_{E,M} \quad (23)$$

And from Eq.(21) we have:

$$\frac{\partial}{\partial M} S(E, M) = \frac{m}{T(E,M)} - \langle (m_3 + m_5)\frac{N-3}{2K_c}\rangle_{E,M} \quad (24)$$

Here $m_3 = \frac{1}{N}\sum_{i=1}^{N} q_i^3$ and $m_5 = \frac{1}{N}\sum_{i=1}^{N} q_i^5$. We define the effective mean field h(E;M) as follows:

$$h(E, M) = -T(E, M)\frac{\partial S(E,M)}{\partial M} \quad (25)$$

using Eq.(24), we have:

$$h(E, M) = -m + T(E, M)\langle (m_3 + m_5)\frac{N-3}{2K_c}\rangle_{E,M} \quad (26)$$

## 5. Results of the Large Deviation Method for the Mean-Field $\varphi^6$ Model

The entropy density $S(\epsilon, m)$ of the mean-field $\varphi^6$ model has been shown as a function of mean energy $\epsilon$ and mean magnetization m for five different values of $\epsilon$ in Fig.1. One of them is above the critical value $\epsilon_c \cong 0.102$ and the rest are below. The effective magnetic field h(E,M) has been plotted as a function of m for different values of $\epsilon$, in Fig.(2). It is worth mentioning that the data in Fig.(1) and Fig.(2) are obtained using the large deviation method i.e., Eq. (15). In Fig.(3) the magnetic susceptibility has been plotted as a function of magnetization m for different values of $\epsilon$.

## 6. Ensemble Inequivalence for Mean Field $\varphi^6$ Model

Now, we study the inequivalence of micro-canonical and canonical ensemble for the case of $\varphi^6$ model which is related to the non-concavity of the entropy $S(\epsilon, m)$. The effective magnetic field h(E;M) is shown in Fig.(2). We observe that h(E;M) can be negative for positive values of m when $\epsilon < \epsilon_c$. In the canonical ensemble, the magnetization $m(\beta, h)$ and the magnetic field h have the same sign. This means that, in the canonical ensemble, the magnetization of the mean field $\varphi^6$ model is always in the direction of the magnetic field. But in the micro-canonical ensemble the situation is different. H(E,M ) and m can have opposite signs, so the canonical and micro-canonical ensembles are inequivalent. For more understanding of this problem in micro-canonical ensemble and its relation with non-concavity of the entropy $S(\epsilon, m)$, we have plotted the magnetic susceptibility and m derivative of the entropy in Figures (3) and (4) respectively. Since T(E;M) is directly proportional to the kinetic energy, it is always positive and this means that the sign of h(E,M) is always opposite to the sign of $\frac{\partial S(\epsilon,m)}{\partial m}$.

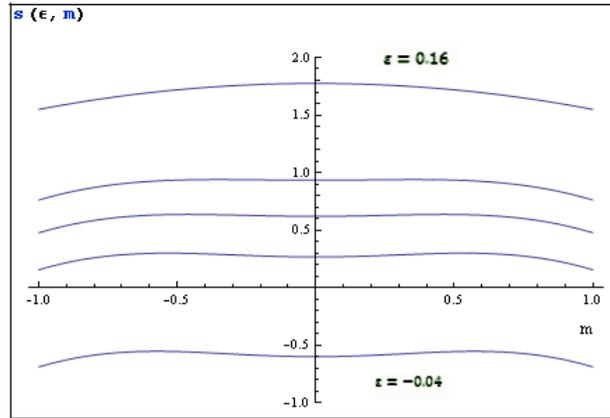

**Figure 1.** Entropy as a function of the mean magnetization m for different values of the mean energy $\epsilon$= 0.16, 0.08, 0.04, 0.00 and -0.04 (from top to bottom)

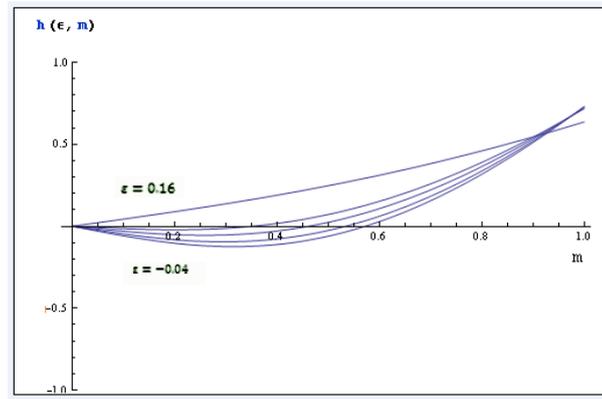

**Figure 2.** Effective magnetic field as a function of magnetization m for different values of $\epsilon = 0.16$, 0.08, 0.04, 0.00 and -0.04 (from top to bottom)

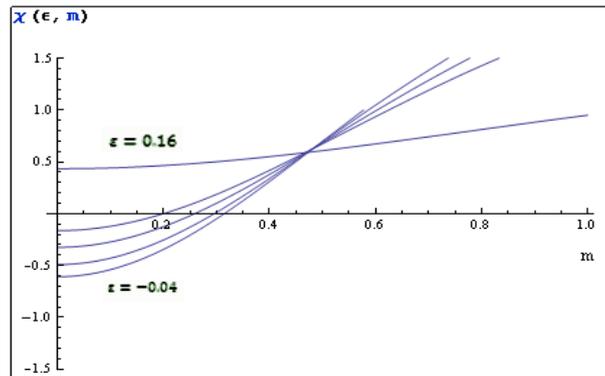

**Figure 3.** Magnetic susceptibility as a function of magnetization m calculated using the large deviation method for different values of $\epsilon$= 0.16, 0.08, 0.04,0.00 and -0.04 (from top to bottom)

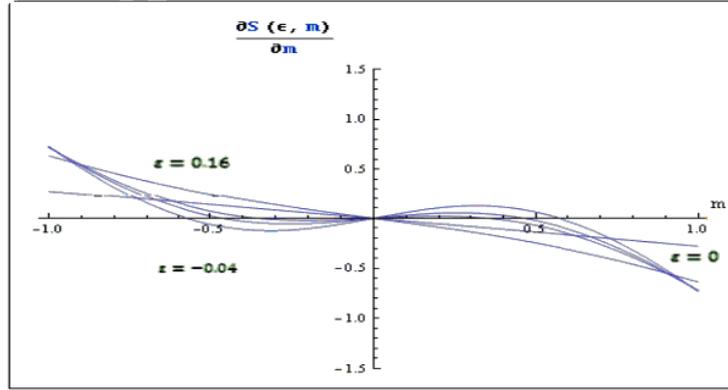

**Figure 4.** $\frac{\partial S(\epsilon,m)}{\partial m}$ for different values of $\epsilon$= 0.16, 0.08, 0.04, 0.00 and -0.04 (from top to bottom), calculated using the large deviation method

Then h(E,M) is negative for m > 0. By comparing Figures (3) and (4), we find that this happens when the entropy $S(\epsilon, m)$ is a concave function of the magnetization e.g., the case for which $\epsilon$= 0.16, then $\frac{\partial S(\epsilon,m)}{\partial m}$ is necessarily negative when m > 0 (Fig.4), which implies that h(E;M) is necessarily positive (for m > 0). This is in agreement with the results of canonical ensemble. Let us now study the magnetic susceptibility in the micro-canonical and canonical ensemble. In the canonical ensemble, the magnetic susceptibility at the constant temperature is defined by the following expression:

$$\chi^T(\beta, h) = \frac{\partial m(\beta,h)}{\partial h} \tag{27}$$

It is easy to show that $\chi^T(\beta, h)$ is always positive. In the micro-canonical ensemble $\chi^T$ is a function of $\epsilon$ and m and given by:

$$\chi^T(E, M) = \left[\frac{\partial h(\epsilon,m)}{\partial m}\bigg|_{T(\epsilon,m)}\right]^{-1} \tag{28}$$

Where $h = \frac{(\frac{\partial S}{\partial m})_\epsilon}{(\frac{\partial S}{\partial \epsilon})_m}$ can be calculated using the first law of thermodynamics $TdS = d\varepsilon + hdm$.

$h(\epsilon, m)$ is given by Eq.(26). Using Eq.(25) and (28) one can show that:

$$\chi^T(\epsilon, m) = -T^{-1}(\epsilon, m) \frac{\partial^2 S}{\partial \epsilon^2} \left[\frac{\partial^2 S}{\partial m^2} \frac{\partial^2 S}{\partial \epsilon^2} - \left(\frac{\partial^2 S}{\partial m \partial \epsilon}\right)^2\right]^{-1} \tag{29}$$

For the mean field $\varphi^6$ model, $\frac{\partial^2 S}{\partial \epsilon^2}$ is always negative, so if for some values of m (when $\epsilon < \epsilon_c$, see fig.3), $\frac{\partial^2 S}{\partial m^2}$ is positive, then $\chi^T$ will be negative as can be checked by Eq.(29). But we already shown that in the canonical ensemble $\chi^T$ is always positive, this again means that micro-canonical and canonical ensembles are nonequivalent. It is worth to mention that this is a consequence of non-concavity of the entropy $S(\epsilon, m)$.

## 7. Conclusion

The entropy of the mean field $\varphi^6$ model is concave as a function of the energy, but is non-concave as a function of the energy and magnetization. This leads to a very important difference between the thermodynamic properties of this model in the micro-canonical and canonical ensembles. We have shown that the effective magnetic field in the micro-canonical ensemble can have a sign opposite to that of the magnetization m, which is in contrast to the case of the canonical ensemble for which the magnetization m is always in the direction of the applied magnetic field. The magnetic susceptibility which in micro-canonical ensemble is a function of the energy and magnetization can be negative but it is always positive in the canonical ensemble. These are two important differences between micro-canonical and canonical ensemble which make them nonequivalent. The mean field $\varphi^6$ model like $\varphi^4$ model displays a first order phase transition due to magnetic field in the canonical ensemble which is a consequence of the non-concavity of the entropy.